\documentclass[pre,twocolumn]{revtex4-2}
\usepackage{amsmath}
\usepackage{hyperref}
\usepackage{graphicx,color}

\begin{document}

\title{Synchronization in a system of Kuramoto oscillators with distributed Gaussian noise}
\author{Alessandro Campa}
\affiliation{National Center for Radiation Protection and Computational Physics, Istituto Superiore di Sanità,
and INFN Roma1, Viale Regina Elena 299, 00161 Roma, Italy}
\author{Shamik Gupta}
\affiliation{Department of Theoretical Physics, Tata Institute of Fundamental Research, Homi Bhabha Road, Mumbai 400005, India}
\begin{abstract}
We consider a system of globally-coupled phase-only oscillators with distributed intrinsic frequencies and evolving in presence of distributed Gaussian, white noise,
namely, a Gaussian, white noise whose strength for every oscillator is a specified function of its intrinsic frequency. In the absence of noise, the model reduces to the celebrated Kuramoto model of spontaneous synchronization. For two specific forms of the mentioned functional dependence and for a symmetric and unimodal distribution
of the intrinsic frequencies, we unveil the rich long-time behavior that the system exhibits, which stands in stark contrast to the case in which the noise strength
is the same for all the oscillators. Namely, in the studied dynamics, the system may exist in either a synchronized or an incoherent or a time-periodic state;
interestingly, all of these states also appear as long-time solutions of the Kuramoto dynamics for the case of bimodal frequency distributions, but in the absence
of any noise in the dynamics.    
\end{abstract}
\maketitle

\section{Introduction}
\label{sec:intro}

The issue of synchronization in systems composed of a large number of similar units characterized by oscillatory
dynamics with distributed intrinsic frequencies is enjoying an increasing interest. This is mainly because of the realization
that one can find many concrete examples of systems of this sort in diverse contexts in Physics, Chemistry, Biology, and even in systems
that can be described within an interdisciplinary approach~\cite{Pikovsky2001}. This ubiquity makes it plausible to guess that the phenomenon of synchronization
is related to some simple properties that are likely to be present in all cases of relevance. In turn, this justifies the study of synchronization through simple models
that capture the relevant features responsible for its emergence in a many-body interacting dynamics.

The model introduced by Kuramoto~\cite{Kuramoto1975} fits perfectly into this strategy of studying the phenomenon of synchronization in a minimalist framework. In fact,
the model represents the individual components as phase-only oscillators. In this way, units that are generally complex nonlinear dissipative systems
are effectively described by just one degree of freedom for every unit, namely, the phase of the individual oscillators, that changes uniformly in time according
to a given frequency $\omega$ before the interaction between the units is turned on. The frequencies of the different oscillators
form a spectrum represented by a distribution $g(\omega)$, and the oscillators are globally coupled with a common coupling constant through the sine of the difference
of phases between the oscillators (see Eq.~\eqref{eq:eom} below, with $D_j=0~\forall~j$). For symmetric and unimodal $g(\omega)$, this relatively simple Kuramoto model
is characterized by a synchronization transition. For small values of the coupling constant, the free uniform motion
of the individual oscillators is not affected significantly by the interaction, and because of the different frequencies, the phases of the individual oscillators get
spread uniformly and independently over $[-\pi, \pi]$ at long times. For sufficiently large values of the coupling constant, the system at long times is characterized
by a macroscopic fraction of the oscillators being locked in frequency to a common frequency, and the system is said to be (partially) synchronized~\cite{Strogatz2000}.
The amount of synchrony increases monotonically with increase in the value of the coupling constant. The system undergoes a phase transition between a low-coupling-strength unsynchronized/incoherent phase and a high-coupling-strength synchronized phase.

Following the aforementioned analysis, investigation of the Kuramoto model has been pursued over the years along several directions: (i) the study of the effects of
replacing the symmetric and unimodal $g(\omega)$ with more general functions~\cite{Kuramoto1984,Martens2009}; (ii) the introduction of noise in
the equations of motion determining the time evolution of the phases~\cite{Pikovsky2001,Sakaguchi1988}; (iii) the introduction of
inertia in the dynamics, transforming the first-order-in-time equations of motion to second-order equations~\cite{Acebron1998,Hong1999,Gupta2014}; (iv) the consideration
of more general interaction functions beyond the simple sine function (although this last point has been comparatively less treated)~\cite{Sakaguchi1987,Daido1992};
(v) the introduction of randomness in the coupling strength, although in a population of oscillators with identical intrinsic
frequencies~\cite{Ko2008a,Ko2008b}.
All these developments have had the purpose of capturing further characteristics of realistic synchronizing systems while remaining in the framework of the Kuramoto model.
The present work may be placed in class (ii) of investigations of the Kuramoto dynamics, albeit with an hitherto-unexplored point of view. Thus,
it is useful to provide a brief summary of the results obtained in previous studies in noisy Kuramoto systems, to see the relationship with the present work.

From the mathematical point of view, the introduction of noise transforms the deterministic equations of motion for the oscillator phases to stochastic
equations. The main motivation for the introduction of noise was to build a model that, while retaining the simplifying feature of describing each unit with only one
degree of freedom (namely, its phase), could nevertheless take into account other relevant features, such as the possible variation of the intrinsic frequencies of the
oscillators~\cite{Sakaguchi1988}. Of course, once introduced, the noise could also be seen, as is usual in statistical mechanics, as a way to model the interaction of the system of oscillators with the external environment. This results in a perturbation of the dynamics of the phases. On physical grounds, it is not surprising that synchronization is more difficult to obtain in a noisy system than in a noiseless system, i.e., given the frequency distribution $g(\omega)$, in order to synchronize the oscillators, the coupling strength has in general to be larger with respect to that which is able to synchronize a noiseless system~\cite{Sakaguchi1988,Gupta2014}. The effect of noise
on a system of Kuramoto oscillators may be studied with the aim to determine the possible change in the nature of asymptotic states as the noise intensity is varied. One may represent the asymptotic states, for any given noise intensity, in a phase diagram, and study how the structure of this phase diagram depends on the noise intensity.
As explained in somewhat more detail later, the changes occurring on varying the noise intensity can depend in turn on the distribution function $g(\omega)$. With a simple unimodal
$g(\omega)$, the noiseless phase diagram is very simple, and it has been described above. Introducing noise and progressively increasing its intensity, the only relevant thing that occurs is the increase of the threshold value of the coupling strength necessary for synchronization~\cite{Sakaguchi1988}. On the other hand, the noiseless phase diagram when $g(\omega)$ is bimodal has a very rich structure~\cite{Martens2009}. By introducing noise and gradually increasing its intensity, the structure changes qualitatively,
tending to become simplified at large noise intensity~\cite{Campa2020}. Another important issue is the influence of the noise on the approach of the oscillator system to the asymptotic state.
This topic lies outside the scope of the paper, but it can be useful to give some details
on this. For noiseless systems, it has been found that the dynamics of the order parameter can be effectively described in some cases in a phase space with a few degrees of freedom, replacing the phase space of the dynamics of all the oscillators~\cite{Ott2008}. By contrast, the introduction of noise forbids this simplifying approach. However, considerable progress has been made also in the direction of obtaining a low dimensional dynamics in spite of the presence of noise~\cite{Tyulkina2018,Goldobin2018,Goldobin2019}.

The above summary concerns systems in which the noise intensity is the same for all the oscillators. However, we stress again
that the reason for introducing noise in the original Kuramoto model was not only to model its interaction
with the external environment, as is usual in statistical mechanics, but also to mimic an inherent variability of the
intrinsic frequencies of the individual oscillators~\cite{Sakaguchi1988}. In fact, the intrinsic frequency of an oscillator is the result of reducing the dynamics of
a complex nonlinear dissipative unit to that of a phase-only oscillator, and, therefore, the intrinsic frequency is in general the combined effect of several internal
dynamical processes. Viewed from this point of view, it is reasonable to assume that different oscillators can be subjected
to different noise levels that model the variability of individual intrinsic frequencies about an average.
As already remarked, in the Kuramoto model and in its extensions, each phase-only oscillator is an effective way to model with just one degree
of freedom a complex unit, i.e., a non-linear dissipative dynamical system generally made of a large number of degrees of freedom. This reduction is based on the
assumption that the dynamics of the unit quickly reaches a limit cycle in course of the dynamical evolution, and subsequently, its dynamics can be described by the evolution of the phase representing the
position along the limit cycle~\cite{Pikovsky2001,Gupta:2018}. The coupling of units described in this way is therefore a coupling between their phases.
Adding a noise of uniform intensity to the original Kuramoto model is a relatively simple procedure, as noted, to describe the variability of the intrinsic frequencies.
If one assumes, as we do in this work, that different oscillators can have different sensitivity to disturbances in their intrinsic frequency (since we consider different
noise levels among the oscillators), in principle, one could consider the possibility that also the coupling strength depends on the particular pair of oscillators
that are coupled. Obviously this would be an even more general model, with a variability of the sensitivity of each oscillator to disturbances in its frequency and a
variability of the sensitivity of the phases to the coupling with other oscillators. We remark that in this case, the relatively simple mean-field approach for the
computation of the order parameter would not be possible (one would probably need to invoke a computation taking into account a sort of quenched disorder for the couplings).
Therefore, in this work, we confine ourselves to the case where only the noise intensity varies among the oscillators. The results show that already in this way, interesting features emerge in the observable properties of the system.

Summarizing, the purpose of this work is to analyze the behavior of Kuramoto oscillators under an inhomogeneous noise that is different for different oscillators.
Thus, we study a
system of oscillators with distributed noise intensity; the aim is to find if, given a frequency distribution $g(\omega)$,
the different intensities of the noise produce qualitatively different results with respect to studies that have considered the uniform-noise case.
We would like to stress that we work in the framework of Gaussian, white, additive noise, although of different intensities, on a system of phase-only
oscillators; a different direction of research would be to take as the basic unit a limit-cycle oscillator subjected to non-Gaussian and coloured
noise~\cite{Goldobin2010}, and considering a population comprising such units that are weakly coupled to one another.
 
As a representative case study, we consider the situation in which a Gaussian, white noise acting on the individual oscillators has a strength that for every oscillator
is a specified function of its intrinsic frequency. In the absence of noise, the model reduces to the Kuramoto model discussed in the foregoing. We consider two distinct
classes of the mentioned functional dependence and for a symmetric and unimodal distribution of the intrinsic frequencies. Our main finding is that the long-time dynamics exhibits a rich behavior, and may exist in either a synchronized or an incoherent or a time-periodic state. Such a behavior is at variance with the case in which the
noise strength is the same for all the oscillators, but which has resemblance with long-time solutions of the Kuramoto dynamics for the case of bimodal frequency
distributions, but in the absence of any noise in the dynamics.    

The paper is structured in the following way. In Section~\ref{sec:model}, we introduce our model of Kuramoto oscillators
under inhomogeneous noise, and give few simple technical features of previous results that have been obtained with the usual
homogeneous noise. Section~\ref{sec:generic-g}, which is the core of our paper, is divided, after an introductory
part, into four subsections. In the first two subsections, we explain how to obtain the bifurcation threshold and the instability threshold
of the incoherent state, while in the third and the fourth subsection, we consider explicitly two different functional forms of the inhomogeneous noise.
Section~\ref{eq:conclusions} discusses the results and draws
some conclusions. Two appendices contain some technical details. 

\section{Model}
\label{sec:model}

Our model of study comprises a system of $N$ globally-coupled phase-only oscillators,
with the phase $\theta_j \in [-\pi,\pi]$ of the $j$-th oscillator evolving in time according
to the dynamics
\begin{align}
\dot{\theta}_{j} & =\omega_{j}+\frac{K}{N}\sum_{k=1}^{N}\sin(\theta_{k}-\theta_{j})+\sqrt{2D_{j}}\eta_{j}(t),
\label{eq:eom}
\end{align}
 where the dot denotes derivative with respect to time, and where $\eta_{j}$ is a Gaussian, white noise with the properties
\begin{align}
\langle\eta_{j}(t)\rangle=0,~~\langle\eta_{j}(t)\eta_{k}(t')\rangle=\delta_{j,k}\delta(t-t').
\end{align}
The angular brackets denote averaging with respect to noise realizations.
In Eq.~\eqref{eq:eom}, $K>0$ is the coupling constant, while the diffusion coefficients $D_{j}$
setting the strength of noise on individual oscillators and the intrinsic frequencies $\omega_{j}$ are both quenched-disordered
random variables sampled independently for every oscillator from given distributions $P(D)$ and $g(\omega)$, respectively.
Note that while we have $\omega_{j}\in(-\infty,\infty),$ we have $D_{j}>0~\forall~j$. The original Kuramoto model
does not include any noise ($D_j=0~\forall~j$), and, correspondingly, the Langevin equations~\eqref{eq:eom} reduce to deterministic equations of motion.

As regards discussing about possible synchronization among the oscillators, one is interested in the dynamical properties and synchronization behavior that the
system exhibits in the long-time limit, or, more practically, after a transient that is quite often of short duration
(the study of the approach to time-asymptotic states is of interest in its own right). This is the framework of our study.
The asymptotic states can sometimes be summarized collectively in a phase diagram, i.e., a diagram with coordinates given by
the relevant dynamical parameters of the system of study. These parameters could be the coupling constant $K$, the noise strength, the
parameters of the frequency distribution $g(\omega)$, etc. In this work, devoted to the behavior of the system~\eqref{eq:eom} under the condition of inhomogeneous noise
($D_j$ different for different oscillators), we have adopted the strategy to present our detailed results obtained
for several representative range of values of these parameters, both analytically and through numerical simulations involving numerical integration of the
dynamics~\eqref{eq:eom}. 
 
\begin{figure}[!htbp]
\includegraphics[width=9cm]{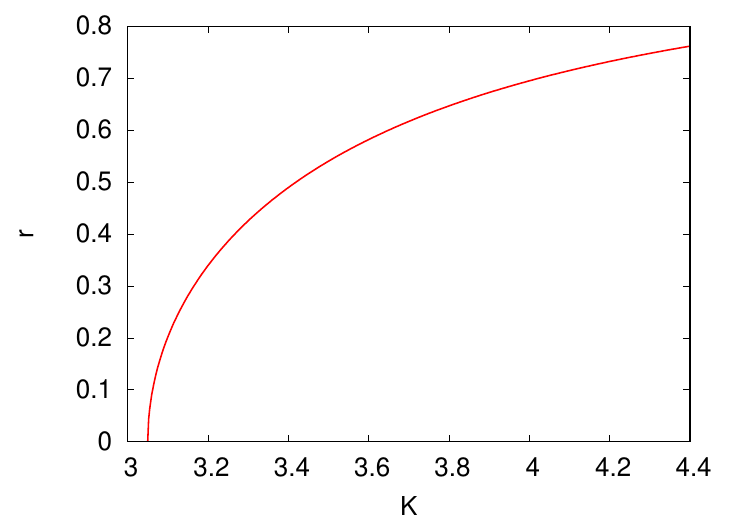}
\caption{For the model~\eqref{eq:eom} with homogeneous noise ($D_j=D=1.0~\forall~j$), the figure shows the stationary order parameter $r$ as a function of $K$. The
frequency distribution is a Gaussian with zero mean and unit variance.}
\label{fig:r-vs-K-bare-model}
\end{figure}

Several limits of the model~\eqref{eq:eom} have been studied in the past~\cite{Gupta:2018}. When the constant $D_j$'s are the same for all the oscillators
(homogeneous-noise case), i.e., $P(D)=\delta(D-D_0)$, with $D_0>0$, we have the following known results. (a) With
$g(\omega)=\delta(\omega-\omega_0)$, choosing a frame rotating uniformly with frequency $\omega_0$ with respect to an
inertial frame, the system can be thought of as being in contact with a heat bath at a temperature $\propto D_0$; it thus
relaxes to a canonical equilibrium stationary state at a temperature $\propto D_0$. (b) With any other form of
$g(\omega)$, we have to distinguish between the simpler case in which this function is symmetric and unimodal with respect to the peak
frequency, and the more general case in which it does not have these properties. In the first case, the system relaxes to a nonequilibrium
stationary state (NESS). In such a stationary state as well as in the canonical equilibrium stationary state achieved with
$g(\omega)=\delta(\omega-\omega_0)$, the system in the limit $N\to \infty$ shows qualitatively similar behavior, namely, a phase transition between
a synchronized and an unsynchronized/incoherent phase as one tunes the value of the coupling constant $K$ at a fixed $D_0$ (or, of $D_0$ at a
fixed $K$). The amount of synchrony in the system is characterized by the synchronization order parameter~\cite{Strogatz2000} 
\begin{align}
	re^{\mathrm{i}\psi}\equiv \frac{1}{N}\sum_{j=1}^N e^{\mathrm{i}\theta_jt};~0\le r \le 1,\psi \in[-\pi,\pi],
	\label{eq:r}
\end{align}
such that $r=0$ (respectively, $r\ne 0$) stands for an unsynchronized (respectively, a synchronized) phase. In this case, the stationary $r$ as a function of
$K$ is zero for $K$ less than a critical value $K_c$, and increases monotonically as a function of $K$ for $K$ larger than this critical
value~\cite{Strogatz2000,Gupta:2018}. A representative plot of $r$ versus $K$, which we will refer back later in the paper, is given in
Fig.~\ref{fig:r-vs-K-bare-model}. The more general case of $g(\omega)$ not being unimodal, in particular when it is bimodal and is the superposition of two symmetric and unimodal distributions centered at two different frequencies, has received overall less attention, but for which several interesting results have been
obtained~\cite{Martens2009,Campa2020}. In this case, in fact,
the long-time state of the system, besides being unsynchronized and synchronized stationary states, can also be a nonstationary periodic state depending on
$K$ and the parameters characterizing the distribution $g(\omega)$. We have to emphasize that
in this work, we refer to nonstationary time-asymptotic states as periodic states using the latter qualification in a loose sense.
In fact, we have no proof of an exact periodicity for the infinite-dimensional system (i.e., in the limit $N \to \infty$, when
the dynamics is described by a Fokker-Planck equation, see later), not even for the particular observable $r(t)$; however, since numerical simulations show quite a
regular and repetitive behavior of $r(t)$ at long times, we take the liberty to refer to such states as periodic.

In the above backdrop, we study in this work the issue of what happens when every oscillator has a different associated noise strength $D_j$ (heterogeneous-noise case),
i.e., when $P(D)$ is not a delta function. In particular, we study the case in which every oscillator has a different intrinsic frequency $\omega_j$ sampled
independently from a given distribution $g(\omega)$ and has also a different noise strength $D_j$, but the randomness in
the latter comes from its explicit dependence on $\omega_j$, i.e., we have
\begin{align}
P(D)=g(\omega)\frac{\mathrm{d}\omega}{\mathrm{d}D},
\end{align}
where the derivative on the right hand side is determined from the given dependence of $D$ on $\omega$, expressed symbolically as $D=D(\omega)$. We may now ask: what
is the nature of the stationary state in such a setting? Distributed $D_{j}$'s imply that formally, we have many heat baths
at different temperatures with all of which the system
is in simultaneous contact, and so the stationary state will be a NESS. However, as remarked in the preceding section, the physical meaning of different
diffusion coefficients $D_j$ is related to the original purpose of the introduction of such coefficients:
they represent both a thermal noise and a variability of the intrinsic frequencies. It is of immediate relevance to address the question: How different
is the structure of the time-asymptotic states from the ones observed when all the $D_{j}$'s are
equal? These are the issues that we take up for a detailed investigation in this work.

\section{Analysis and results}
\label{sec:generic-g}

In the limit $N\to \infty$, our system of study~\eqref{eq:eom} may be characterized by a single-oscillator distribution function
$f(\theta,\omega,t)$, defined such that among all oscillators with  the same intrinsic frequency $\omega$ (and
consequently, same noise parameter $D=D(\omega)$), the quantity $f(\theta,\omega,t)\mathrm{d}\theta$ gives the fraction
that have their phase in $[\theta,\theta+\mathrm{d}\theta]$ at time $t$. This function is normalized as
\begin{align}
	\int_{-\pi}^{\pi} \mathrm{d}\theta~f(\theta,\omega,t)=1~\forall~\omega,t,
	\label{eq:norm1}
\end{align}
and is also periodic: $f(\theta+2\pi,\omega,t)=f(\theta,\omega,t)$. We recall that in the usual case when the diffusion coefficient has the same value $D$ for
all oscillators, the function $f(\theta,\omega,t)$ evolves in time according to the Fokker-Planck equation~\cite{Gupta:2018}
\begin{align}
	\frac{\partial f}{\partial t}=-\frac{\partial }{\partial \theta}\left[(\omega+Kr\sin(\psi-\theta))f\right]
	+D\frac{\partial^2 f}{\partial \theta^2},
	\label{eq:Fokker-Planck}
\end{align}
with
\begin{align}
	re^{\mathrm{i}\psi}=\int \mathrm{d}\theta \mathrm{d}\omega~e^{\mathrm{i}\theta}f(\theta,\omega,t)g(\omega).
\end{align}

In a stationary state, when $r$ and $\psi$ become time independent, measuring all $\theta$'s with respect to the
stationary $\psi$ leads to the stationary distribution $f_\mathrm{st}(\theta,\omega)$ satisfying 
\begin{align}
	&0=-\frac{\partial }{\partial \theta}\left[(\omega-Kr\sin \theta)f_\mathrm{st}\right]
	+D\frac{\partial^2 f_\mathrm{st}}{\partial \theta^2},\label{eq:steady1-1}\\
	&r=\int \mathrm{d}\theta \mathrm{d}\omega~\cos \theta~f_\mathrm{st}(\theta,\omega)g(\omega),\label{eq:steady2-1}\\
	&0=\int \mathrm{d}\theta \mathrm{d}\omega~\sin \theta~f_\mathrm{st}(\theta,\omega)g(\omega). \label{eq:steady2-2}
\end{align}
Equation~\eqref{eq:steady1-1} has the solution~\cite{Gupta:2018} 
\begin{align}
	&f_\mathrm{st}(\theta,\omega)=Ce^{\frac{Kr\cos \theta+\omega \theta}{D}}\nonumber \\
	&\times \left[1+\left(e^{-\frac{2\pi \omega}{D}}-1\right)\frac{\int_{0}^\theta \mathrm{d}\theta'
	e^{-\frac{Kr\cos \theta'+\omega \theta'}{D}}}{\int_{0}^{2\pi} \mathrm{d}\theta'
	e^{-\frac{Kr\cos \theta'+\omega \theta'}{D}}}\right],
	\label{eq:solution}
\end{align}
with $C=f_\mathrm{st}(0,\omega)e^{-Kr/D}$ being the normalization constant to fix the condition~\eqref{eq:norm1}, and in
which $r$ is to be determined self-consistently by inserting Eq.~\eqref{eq:solution} in Eq.~\eqref{eq:steady2-1}. Note that
$f_\mathrm{st}(\theta,\omega,D)=f_\mathrm{st}(-\theta,-\omega,D)$ automatically satisfies Eq.~\eqref{eq:steady2-2}.

One of the main purposes of this work is to show that the inhomogeneity of the diffusion coefficient $D$ among the different oscillators can give rise to a richness of possibilities in the long-time state of the dynamics~\eqref{eq:eom}.
In fact, we will show that with an inhomogeneous diffusion coefficient and a unimodal $g(\omega)$,
the behavior can be, in some cases, similar to the noiseless system with a bimodal $g(\omega)$.
To unveil this fact, we take $g(\omega)$ to be a standard Gaussian with zero mean and unit variance in the rest of the paper.

To proceed with our analysis with the inhomogeneous-$D$ case, we note as a first thing that the Fokker-Planck
equation~\eqref{eq:Fokker-Planck} and hence the equation~\eqref{eq:solution} for the stationary solution have the same form even when $D$ is
$\omega$-dependent, $D=D(\omega)$. We will use this fact in our subsequent analysis, in which we will study two different and representative forms of the
function $D(\omega)$. We reproduce here the equations for convenience:
\begin{align}
	\frac{\partial f}{\partial t}=-\frac{\partial }{\partial \theta}\left[(\omega+Kr\sin(\psi-\theta))f\right]
	+D(\omega)\frac{\partial^2 f}{\partial \theta^2},
	\label{eq:Fokker-Planck-D}
\end{align}
and 
\begin{align}
	&f_\mathrm{st}(\theta,\omega)=Ce^{\frac{Kr\cos \theta+\omega \theta}{D(\omega)}}\nonumber \\
	&\times \left[1+\left(e^{-\frac{2\pi \omega}{D(\omega)}}-1\right)\frac{\int_{0}^\theta \mathrm{d}\theta'
	e^{-\frac{Kr\cos \theta'+\omega \theta'}{D(\omega)}}}{\int_{0}^{2\pi} \mathrm{d}\theta'
	e^{-\frac{Kr\cos \theta'+\omega \theta'}{D(\omega)}}}\right].
	\label{eq:solution-D}
\end{align}
We will study one case in which $D(\omega)$ is a simple two-valued function, and also a case in which $D(\omega)$ is a smooth function. However, before specializing
to these choices of $D(\omega)$, we derive some general expressions that will be used in our analysis. We will in the following two subsections derive (i) the bifurcation threshold $K_c$, namely, the value of $K$ where a synchronized stationary state, a solution with positive $r$ of Eq.~\eqref{eq:Fokker-Planck-D},
bifurcates from the incoherent stationary state with $r=0$ (see, e.g., the beginning of the curve at $r=0$ in
Fig. \ref{fig:r-vs-K-bare-model}), and (ii) the instability threshold $K_i$, namely, the value of $K$ beyond which the incoherent stationary state is no longer linearly
stable under evolution according to Eq.~\eqref{eq:Fokker-Planck-D}. 

\subsection{The bifurcation threshold}
\label{sec:bifurcation-threshold}

From Eqs.~\eqref{eq:Fokker-Planck-D} and~\eqref{eq:solution-D}, we see that $r=0$, and correspondingly,
$f_\mathrm{st}=1/(2\pi)$, is always an acceptable stationary solution for all values of $K$. Thus, the incoherent state is always a stationary solution of the
dynamics~\eqref{eq:eom} for all $K$, but its stability is an issue that has to be checked, which we will address shortly. As with the homogeneous-noise case, it
will so turn out that the incoherent state is unstable beyond a critical value of $K$. 

Equations~\eqref{eq:Fokker-Planck-D} and~\eqref{eq:solution-D} can also admit a nonzero stationary solution $r>0$, corresponding to a synchronized state, and we want
to now study when as a function of $K$ such a nonzero solution bifurcates from the $r=0$ stationary solution. To find the value $K_c$ of $K$ at which such
a bifurcation occurs, we substitute Eq.~\eqref{eq:solution-D} into Eq.~\eqref{eq:steady2-1} and expand the right hand side of the latter equation as a power
series in $r$ in the limit $r\to 0$. Keeping only the first two terms of the series, one arrives at the
following expression~\cite{Gupta:2018}:
\begin{widetext}
\begin{eqnarray}
\label{powerself}
r=\frac{Kr}{2}\!\!\!\int \mathrm{d} \omega~g(\omega) \frac{D(\omega)}{\omega^2 +
(D(\omega))^2}-\frac{K^3r^3}{4}\!\!\!\int \mathrm{d}\omega~ \frac{g(\omega)}{D(\omega)} \left[\frac{1}{\omega^2 +4(D(\omega))^2}
-\frac{\omega^2}{\left(\omega^2 + (D(\omega))^2\right)^2}\right].
\end{eqnarray}
\end{widetext}
The desired bifurcation threshold $K_c$ is the
value of $K$ for which the coefficient of $r$ on the right hand side of
Eq.~\eqref{powerself} is equal to $1$. We thus get
\begin{equation}
\label{kthreshold}
K_c = 2 \left[ \int \mathrm{d}\omega~ g(\omega) \frac{D(\omega)}{\omega^2 + (D(\omega))^2} \right]^{-1}.
\end{equation}

\subsection{The instability threshold}
\label{sec:instability-threshold}

As promised, we now perform the linear stability analysis of the incoherent stationary state
$f_\mathrm{st}(\theta,\omega) = 1/(2\pi)$. To this end, expanding
$f(\theta,\omega,t)$ as
\begin{equation}
\label{expandf}
f(\theta,\omega,t) = \frac{1}{2\pi} + \delta f(\theta,\omega,t),
\end{equation}
with $\delta f \ll 1$, we see on account of the normalization~\eqref{eq:norm1} that $\int_{-\pi}^{\pi}
\mathrm{d}\theta~\delta f(\theta,\omega,t)=0~\forall~\omega,t$. Substituting the above expansion in
Eq.~\eqref{eq:Fokker-Planck-D}, and keeping only terms that are at most linear in $\delta f$, we obtain
the linear equation
\begin{eqnarray}
\label{FPone_kura_linear}
&&\frac{ \partial}{\partial t}\delta f(\theta,\omega,t) = - \omega \frac{\partial}{\partial \theta}\delta f(\theta,\omega,t)
+ D(\omega) \frac{\partial^2}{\partial \theta^2}\delta f(\theta,\omega,t)
\nonumber 
\\
&&+\frac{K}{2\pi} \int \mathrm{d}\omega' g(\omega') \int \mathrm{d} \theta' 
\cos(\theta - \theta')\delta f(\theta',\omega',t). 
\end{eqnarray}
Effecting a Fourier expansion:
$\delta f(\theta,\omega,t) = \sum_{k=-\infty}^{+\infty} \widehat{\delta f}_k(\omega,t) e^{\mathrm{i} k \theta}$, 
and then substituting in Eq.~\eqref{FPone_kura_linear}, we get for the $k$-th Fourier component that
\begin{eqnarray}
\label{four_deltaf_k}
&&\frac{ \partial}{\partial t}\widehat{\delta f}_k(\omega,t) = - \mathrm{i} k \omega \widehat{\delta f}_k(\omega,t)
- D(\omega) k^2 \widehat{\delta f}_k(\omega,t) \nonumber \\
&&+\frac{K}{2}\left( \delta_{k,1}+\delta_{k,-1}\right) \int \mathrm{d} \omega' g(\omega') 
\widehat{\delta f}_k(\omega',t).
\end{eqnarray}
For $k\ne \pm 1$, one has an exponential decay of
$\widehat{\delta f}_k(\omega,t)$ with time, with rate equal to $D(\omega)k^2$. 
For $k = \pm 1$, using
\begin{equation}
\label{autovdeltaf}
\widehat{\delta f}_{\pm 1}(\omega,t) = \widetilde{\delta f}_{\pm 1}(\omega, \lambda) e^{\lambda t}
\end{equation}
 in Eq.~\eqref{four_deltaf_k} yields
\begin{equation}
\label{four_deltaf_kpm1}
\left(\lambda \pm \mathrm{i} \omega + D(\omega)\right) \widetilde{\delta f}_{\pm 1}(\omega,\lambda)
= \frac{K}{2} \int \mathrm{d} \omega' g(\omega') 
\widetilde{\delta f}_{\pm 1}(\omega',\lambda).
\end{equation}
It then follows that one has a continuous spectrum of stable modes, one for each $\omega$ value in the support of $g(\omega)$; If $\omega_0$
is one such value, the corresponding stable mode for $\widetilde{\delta f}_{\pm 1}$ has
$\lambda = -D(\omega_0) \mp \mathrm{i} \omega_0$~\cite{Gupta:2018}. On the other hand, there is also a discrete spectrum, with the corresponding $\lambda$ found by
rewriting Eq.~\eqref{four_deltaf_kpm1} as
\begin{equation}
\label{sol_deltaf_kpm1_disc}
\widetilde{\delta f}_{\pm 1}(\omega,\lambda) = \frac{K}{2\left(\lambda \pm \mathrm{i} \omega + D(\omega)\right)}
\int \mathrm{d} \omega'
g(\omega') \widetilde{\delta f}_{\pm 1}(\omega',\lambda).
\end{equation}
Multiplying both sides by $g(\omega)$, integrating with respect to $\omega$ and using $\int \mathrm{d} \omega'
g(\omega') \widetilde{\delta f}_{\pm 1}(\omega',\lambda)\ne 0$~\cite{note1}, we obtain the
dispersion relation
\begin{equation}
\label{disper_rel}
\frac{K}{2} \int \mathrm{d} \omega \frac{g(\omega)}{\lambda \pm \mathrm{i} \omega + D(\omega)} = 1.
\end{equation}
If $\lambda_\mathrm{r}$ and $\lambda_\mathrm{i}$ are the
real and the imaginary part of $\lambda$, then the real and the imaginary part of Eq.~\eqref{disper_rel} give
\begin{eqnarray}
\label{disper_rel_real}
&&\frac{K}{2}\int \mathrm{d} \omega~g(\omega) \frac{\lambda_\mathrm{r} + D(\omega)}
{\left(\lambda_\mathrm{r} + D(\omega)\right)^2 + \left(\lambda_\mathrm{i} \pm \omega \right)^2} = 1, \\
\label{disper_rel_imag}
&&\int \mathrm{d} \omega~g(\omega) \frac{\lambda_\mathrm{i} \pm \omega}
{\left(\lambda_\mathrm{r} + D(\omega)\right)^2 + \left(\lambda_\mathrm{i} \pm \omega \right)^2}= 0.
\end{eqnarray}

For the usual case of an $\omega$-independent $D$ and a symmetric and unimodal $g(\omega)$ centered at $\omega=0$,
one can prove~\cite{Gupta:2018} that Eq.~\eqref{disper_rel_imag} can be satisfied only for real values of $\lambda$, and, furthermore,
for $\lambda_\mathrm{i}=0$, it is trivially satisfied. Therefore, one is left with the search of the solution
$\lambda_\mathrm{r}$ of Eq.~\eqref{disper_rel_real} after posing $\lambda_\mathrm{i}=0$ in it. The instability threshold $K_i$ of the incoherent state is obtained
by further posing $\lambda_\mathrm{r}=0$. Following this procedure, one finds that in this case, we have $K_i = K_c$~\cite{Gupta:2018}. This simple situation does
not arise when $g(\omega)$ is not symmetric and unimodal. However, and this is what
concerns us in this work, when $D(\omega)$ is $\omega$-dependent, this coincidence of $K_i$ with $K_c$ is not verified also with a symmetric and unimodal $g(\omega)$
such as the standard Gaussian we are considering. Then, one can proceed with the following strategy. It is still the case that the instability threshold $K_i$ is
characterized by being a solution of
Eqs.~\eqref{disper_rel_real} and \eqref{disper_rel_imag} for $\lambda_\mathrm{r}=0$. Since $K$ does not appear in the second
equation, one can numerically look for its solution $\lambda_\mathrm{i}$ after posing $\lambda_\mathrm{r}=0$; then plugging
this value of $\lambda_\mathrm{i}$ in the first equation with $\lambda_\mathrm{r}=0$, one obtains directly the value of
$K_i$.

\subsection{The choice of $D=D(\omega)$ as a step function}
\label{sec:choice1}

Here, we consider the choice of $D=D(\omega)$ as a step function, namely, $D=D_1$ for $|\omega|< \omega_0$
and $D=D_2$ for $|\omega|> \omega_0$, where $D_1$, $D_2$ and $\omega_0$ are given constants. This choice will prove
useful in having an analytical characterization of the consequences of an inhomogeneous diffusion coefficient and will allow for the bifurcation threshold
associated with the synchronization transition
to be analytically derived.

Let us then begin with the determination of the bifurcation threshold $K_c$. We have to insert our step function
$D(\omega)$ given above in Eq.~\eqref{kthreshold}, obtaining, after writing explicitly the standard Gaussian for $g(\omega)$, that
\begin{eqnarray}
\label{kthreshold_step}
K_c &=& \sqrt{8\pi}\left[ \int_{|\omega|\le \omega_0} \mathrm{d}\omega~ e^{-\frac{\omega^2}{2}} \frac{D_1}{\omega^2 + D_1^2}
\right. \nonumber \\
&+& \left. \int_{|\omega|\ge \omega_0}  \mathrm{d}\omega~ e^{-\frac{\omega^2}{2}} \frac{D_2}{\omega^2 + D_2^2} \right]^{-1}.
\end{eqnarray}
The computation of the integrals is given in Appendix~\ref{sec:app1}, and we get
\begin{widetext}
\begin{eqnarray}
\label{kbiffin-main}
K_c
&=& \sqrt{\frac{8}{\pi}} \left[ \mathrm{erfc}\left(\frac{D_1}{\sqrt{2}}\right)e^{\frac{D_1^2}{2}}
-\frac{2}{\pi} \int_{\arctan \left(\frac{\omega_0}{D_1}\right)}^{\frac{\pi}{2}} \mathrm{d} \phi \, e^{-\frac{D_1^2}{2}\tan^2\phi}
+\frac{2}{\pi} \int_{\arctan \left(\frac{\omega_0}{D_2}\right)}^{\frac{\pi}{2}} \mathrm{d} \phi \, e^{-\frac{D_2^2}{2}\tan^2\phi}
\right]^{-1},
\end{eqnarray}
\end{widetext}
where $\mathrm{erfc}(x)$ is the complementary error function.

\begin{figure}[!htbp]
\includegraphics[width=9cm]{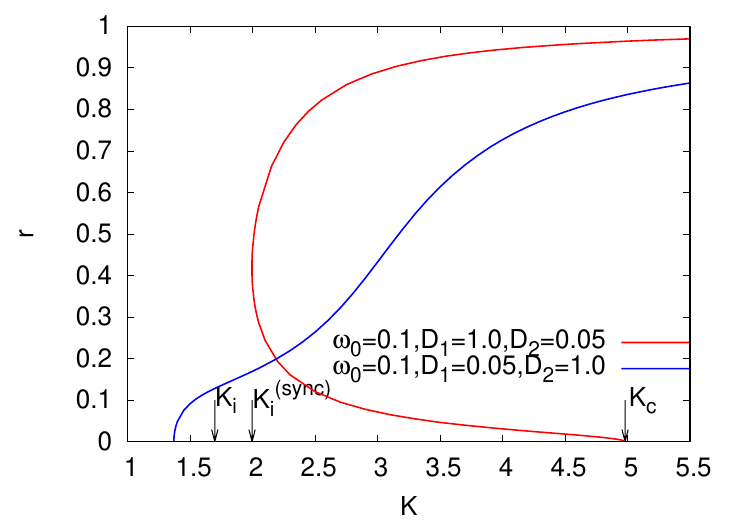}
\caption{For the model~\eqref{eq:eom}, the figure shows for the case studied in Section~\ref{sec:choice1} of the particular dependence of the diffusion coefficient
$D$ on $\omega$ the numerically-computed order parameter $r$ as a function $K$. The frequency distribution is a Gaussian with zero mean and unit variance. One may
observe for case (i): $\omega_0=0.1, D_1=1.0, D_2=0.05$ (respectively, for case (ii): $\omega_0=0.1, D_1=0.05,D_2=1.0$) a subcritical (respectively, a supercritical)
bifurcation of the synchronized state from the incoherent state (the tangent of the curve at $K_c$ for case (i) is vertical as it is for case (ii), although the
scale of the figure does not allow to have a clear visualization of this fact for case (i)). The results are obtained by solving numerically the self-consistent
equation for $r$ as detailed in the text. Here, $K_i$ and $K_i^{(\mathrm{sync})}$ are the instability threshold, respectively, of the incoherent and the synchronized
state. On the other hand, $K_c$ is the bifurcation threshold between the incoherent and the synchronized state. For case (i), these values are explicitly
denoted with arrows, and the figure shows that $K_i \ne K_i^{(\mathrm{sync})}\ne K_c$; for case (ii), we have
$K_i=K_i^{(\mathrm{sync})}=K_c$. For case (i), $K_i^{(\mathrm{sync})}$ is given by the leftmost $K$ value of the curve.}
\label{fig:CaseA-r-vs-K}
\end{figure}

\begin{figure}[!htbp]
\includegraphics[width=9cm]{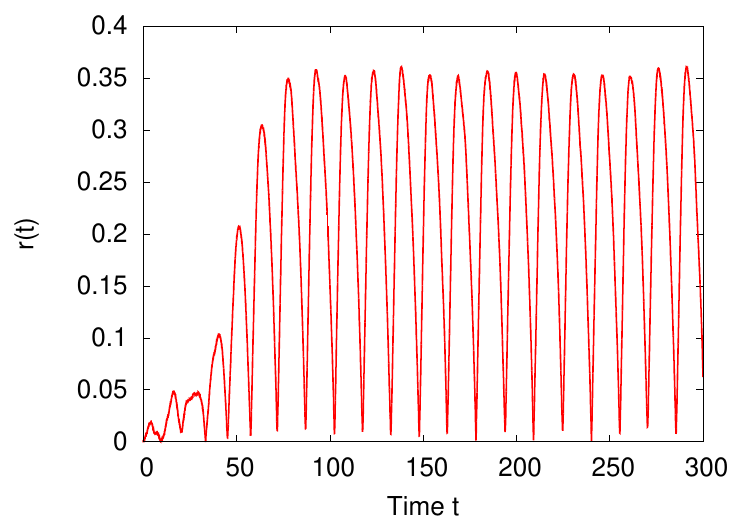}
\includegraphics[width=9cm]{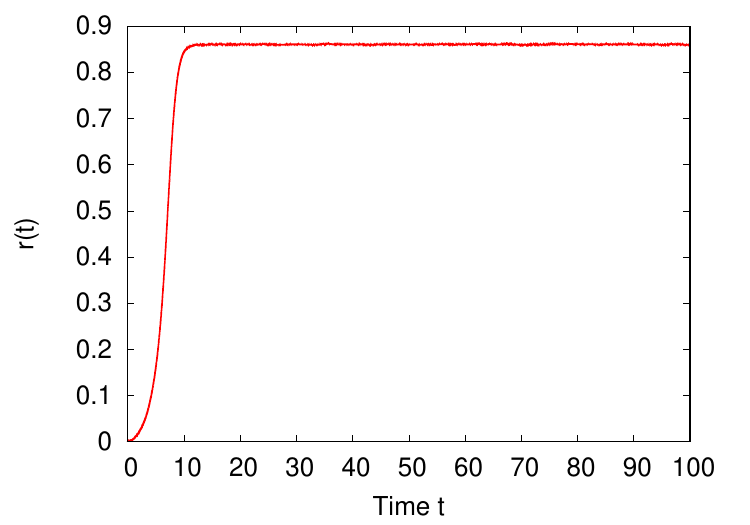}
\caption{For the model~\eqref{eq:eom}, the figure shows for the case (i) studied in Section~\ref{sec:choice1} of the particular dependence of the diffusion
coefficient $D$ on $\omega$ the order parameter $r$ versus time and for two values of $K$: Upper panel: $K=1.88$ is chosen to be between $K_i$ and
$K_i^{(\mathrm{sync})}$, in which case the dynamics relaxes to a time-periodic state. Lower panel: $K=2.75$ is chosen to be larger than $K_i^{(\mathrm{sync})}$,
in which case the dynamics relaxes to a synchronized stationary state. The frequency distribution is a Gaussian with zero mean and unit variance. For both panels, the initial state has the $\theta$'s uniformly and independently distributed
in $[-\pi,\pi]$, while the number of oscillators is $N=10^5$. The data
are obtained from numerical integration of the equations of motion~\eqref{eq:eom}.}
\label{fig:simulation-CaseA}
\end{figure}

With the above background, we now present numerical results on $r$ as a function $K$ for representative values of the parameters $\omega_0, D_1, D_2$. We will
establish consistency of our results with those obtained from numerical integration of the dynamics~\eqref{eq:eom}. For numerical computation of $r$ versus $K$,
we employ the method discussed in Ref.~\cite{Gupta:2018} and summarized in Appendix~\ref{sec:app2}. For parameter values, we consider two representative
cases: (i) $\omega_0=0.1, D_1=1.0, D_2=0.05$, and (ii) $\omega_0=0.1, D_1=0.05, D_2=1.0$. Clearly, these two choices complement each other. While in the first choice,
low-frequency oscillators have larger noise in their dynamics than the high-frequency oscillators, the reverse is true for the second choice. For case (i), using
the method detailed in the preceding subsection, one obtains the bifurcation threshold and the instability threshold given respectively by
$K_c \approx 4.978, K_i \approx 1.697$. For the second choice, we have instead that $K_c=K_i \approx 1.371$. The fact that
$K_c$ and $K_i$ are unequal in one case and equal in the other case already hints at the system behaving very differently
in as far as the long-time behavior of the order parameter $r$ as a function of $K$ is concerned. In
Fig.~\ref{fig:CaseA-r-vs-K}, we show the numerically-computed $r$ as a function of $K$ for the aforementioned two choices of the parameters $D_1$ and $D_2$. We
observe for case (i) of decreasing $D$ with $\omega$ (respectively, for case (ii) of increasing $D$ with $\omega$) a subcritical (respectively, a supercritical)
bifurcation of the synchronized state from the incoherent state. The thresholds $K_c$ and $K_i$ for case (i) are explicitly indicated with arrows in the figure,
unlike for case (ii), for which both these thresholds are at the beginning of the blue line on the $K$ axis. 
We have checked that the values of the bifurcation threshold $K_c$ for the two cases, obtained from the
curves $r(K)$ shown in the figure, match with the aforementioned theoretical values for these quantities. Moreover, for
case (i), we have indicated another relevant value of $K$, i.e. $K_i^{(\mathrm{sync})} \approx 1.995$, which is defined as the leftmost $K$ value of the curve $r(K)$.
This value is nothing but the instability threshold of the synchronized state
(this state exists only for $K\ge K_i^{(\mathrm{sync})}$). We thus see that there is a range of $K$ values between the instability thresholds $K_i \approx 1.697$
and $K_i^{(\mathrm{sync})} \approx 1.995$ of the incoherent and the synchronized stationary state, respectively, in which the system does not exhibit any stationary
state at long times. We note that case (ii) of increasing $D$ with $\omega$ shows a behavior of $r$ versus $K$ that is not very different from the behavior observed
with the homogeneous-noise case, see Fig.~\ref{fig:r-vs-K-bare-model}. For
case (ii), one has $K_i=K_i^{(\mathrm{sync})}=K_c$, and the system always has a well-defined stationary state at long times. The plot of
Fig.~\ref{fig:simulation-CaseA}, showing the order parameter $r$ as a function of time from numerical integration of the dynamics~\eqref{eq:eom} for case (i) above,
is consistent with the predictions of Fig.~\ref{fig:CaseA-r-vs-K}. Namely, for $K$ between $K_i$ and $K_i^{(\mathrm{sync})}$, the system relaxes to a nonstationary
state (which is a time-periodic state, as seen in Fig.~\ref{fig:simulation-CaseA}(a)). On the other hand, for $K>K_i^{(\mathrm{sync})}$, the system relaxes to a
synchronized stationary state. For $K<K_i$, the system relaxes to an incoherent stationary state (data not shown).

In the light of the above, one may wonder whether the non-existence of a stationary state and the observed subcritical bifurcation for case (i), results that are very
different from those for the homogeneous-noise case, are a consequence of the underlying $D$ versus $\omega$ dependence being a step and not a smooth function. To explore further this point, we are thus led in the following subsection to study the case in which $D(\omega)$ decreases as a function of $\omega$, but does so in the form of
a smooth functional dependence. 

\subsection{The choice $D(\omega)=1-\gamma|\omega|/\sqrt{1+\gamma^2\omega^2}$}
\label{sec:choice2}

Here, we consider $D=D(\omega)$ to be monotonically decreasing as a function of $|\omega|$, so that oscillators with large intrinsic frequencies are the ones
with small noise strengths. We choose to study the behavior of our system when the function $D(\omega)$ has the following form:
\begin{equation}
D(\omega) = 1 - \frac{\gamma |\omega|}{\sqrt{1+\gamma^2\omega^2}} \, .
\label{dsmooth}
\end{equation}
The parameter $\gamma>0$ sets the rate of decrease of $D(\omega)$ with increasing $|\omega|$. Apart from the cusp in
$\omega=0$, the above is a smooth function for all values of $\omega \ne 0$.

\begin{figure*}[!htbp]
 \begin{minipage}{0.5\linewidth}
  \centering
  \includegraphics[width=9cm]{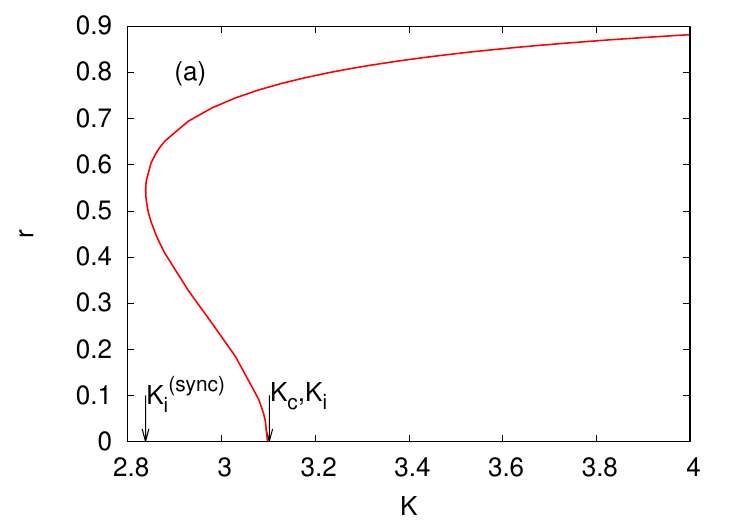}
 \end{minipage}%
 \begin{minipage}{0.5\linewidth}
  \centering
  \includegraphics[width=9cm]{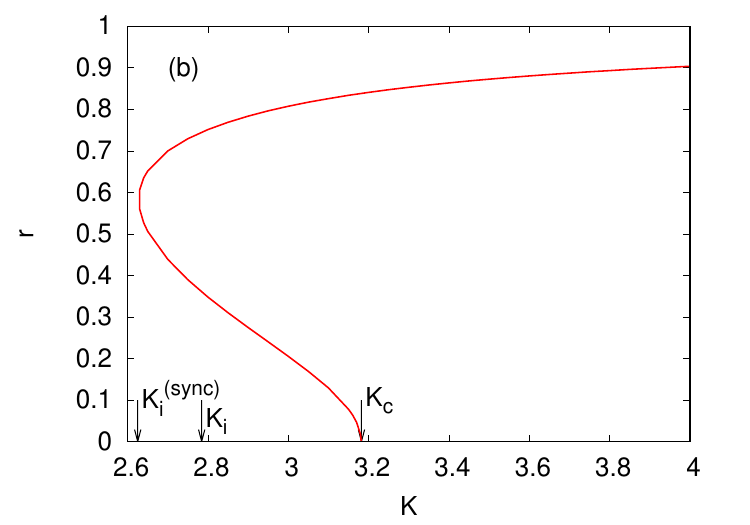}
  \end{minipage}
  \begin{minipage}{0.5\linewidth}
  \centering
  \includegraphics[width=9cm]{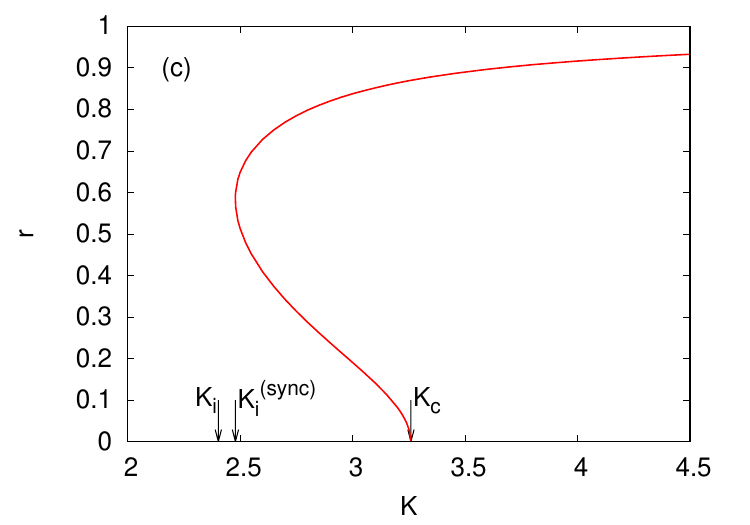}
 \end{minipage}%
 \begin{minipage}{0.5\linewidth}
  \centering
  \includegraphics[width=9cm]{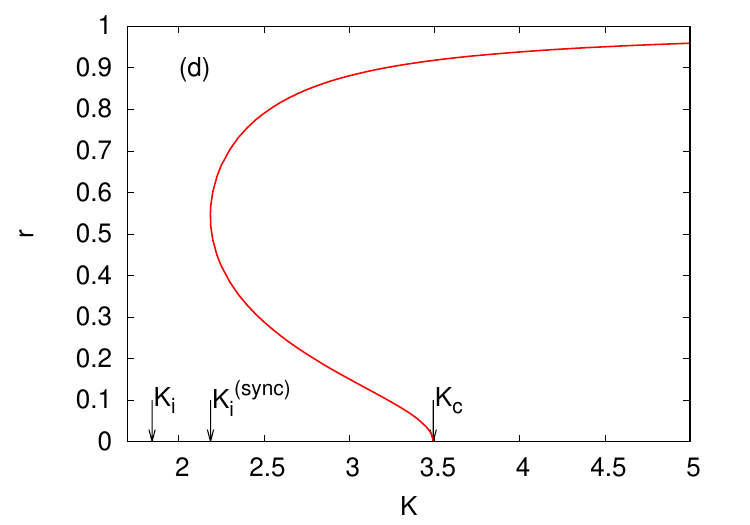}
  \end{minipage}
  \caption{For the model~\eqref{eq:eom}, the figure shows for the case studied in Section~\ref{sec:choice2} of the particular dependence of the diffusion
	coefficient $D$ on $\omega$ the numerically-computed order parameter $r$ as a function of $K$. The values of the parameter $\gamma$ are
	$\gamma=1.0$ [panel (a)], $\gamma=1.5$ [panel (b)], $\gamma=2.0$ [panel (c)], and
	$\gamma=4.0$ [panel (d)]. The frequency distribution is a Gaussian with zero mean and unit variance. The instability
	thresholds $K_i$ and $K_i^{(\mathrm{sync})}$ and the bifurcation threshold $K_c$ are shown in the figure.}
\label{fig:CaseB-r-vs-K}
 \end{figure*}

In Fig.~\ref{fig:CaseB-r-vs-K}, we show the plots of $r(K)$ for four representative values of $\gamma$. We now discuss
the implications of these curves. Let us begin with the two large values of $\gamma$, i.e., $\gamma=2.0$ and $\gamma=4.0$,
for which the corresponding curves $r(K)$ are given, respectively, in panels (c) and (d) of Fig.~\ref{fig:CaseB-r-vs-K}.
A comparison with the red curve of Fig.~\ref{fig:CaseA-r-vs-K} shows that the qualitative situation is the same as that
of case (i) of Section~\ref{sec:choice1}: there is
a subcritical bifurcation of the synchronized state from the incoherent state at $K=K_c$; furthermore, the instability thresholds $K_i$
of the incoherent stationary state and $K_i^{(\mathrm{sync})}$ of the synchronized stationary state
are in the order $K_i < K_i^{(\mathrm{sync})} < K_c$. This demonstrates that for smooth $D(\omega)$, we find the
same qualitative behavior as was reported in the preceding subsection for $D(\omega)$ given as a step function. By increasing $K$ beyond $K_i$, the incoherent
stationary state becomes unstable. However, since for $K_i < K < K_i^{(\mathrm{sync})}$, there is no stable stationary state, the
system settles into a periodic time-asymptotic state in this range of $K$ values. We do not show plots of $r(t)$ as in Fig.~\ref{fig:simulation-CaseA}, since these
plots would present the same feature, i.e., a nonstationary time-asymptotic periodic state for
a value of $K$ between $K_i$ and $K_i^{(\mathrm{sync})}$, and a stationary synchronized state for a value of
$K$ larger than $K_i^{(\mathrm{sync})}$. For reference, we give the values of the various thresholds. For $\gamma = 2.0$,
we have $K_i \approx 2.403$, $K_i^{(\mathrm{sync})} \approx 2.479$, and $K_c \approx 3.259$, while for $\gamma = 4.0$,
the values are $K_i \approx 1.843$, $K_i^{(\mathrm{sync})} \approx 2.186$, and $K_c \approx 3.493$.
Our results indicate that for these two cases, as well for case (i) of Section \ref{sec:choice1}, the value $K_i$ marks a
Hopf bifurcation. 

We now move on to discuss the other cases in Fig.~\ref{fig:CaseB-r-vs-K}, which correspond to smaller values of $\gamma$, and which present different and interesting
characteristics. For $\gamma =1.0$, we see from panel (a) of Fig.~\ref{fig:CaseB-r-vs-K} that the instability threshold
$K_i$ of the incoherent state coincides with the bifurcation threshold $K_c$ (we have $K_c = K_i \approx 3.099$). However, since
this bifurcation is
subcritical, and the instability threshold $K_i^{(\mathrm{sync})}$ is smaller than $K_i$
($K_i^{(\mathrm{sync})} \approx 2.839$), one is led to deduce that
we have now a first-order transition between the incoherent and the synchronized stationary state, with an associated
hysteresis behavior. This means that we have a case of bistability: in the range
$K_i^{(\mathrm{sync})} < K < K_i$, depending on the initial condition, the system can settle asymptotically
into either an incoherent or a synchronized state. On the other hand, by adiabatically tuning the coupling constant $K$,
the system with increasing $K$ remains in the incoherent state all the way up to $K=K_i$, when it jumps to the synchronized state;
following the hysteresis cycle, but now with decreasing $K$, the system remains in the synchronized state all the way up to $K=K_i^{(\mathrm{sync})}$,
when it jumps back to the incoherent state. This picture has been confirmed by our numerical integration results.

\begin{figure}[!htbp]
\includegraphics[width=9cm]{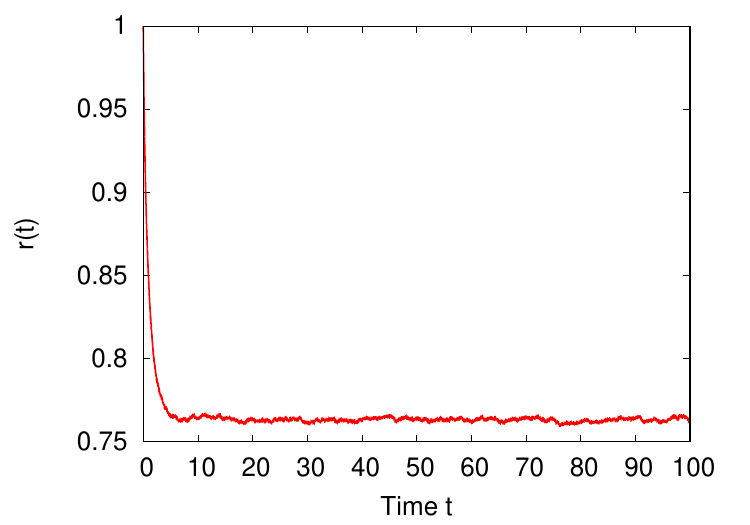}
\includegraphics[width=9cm]{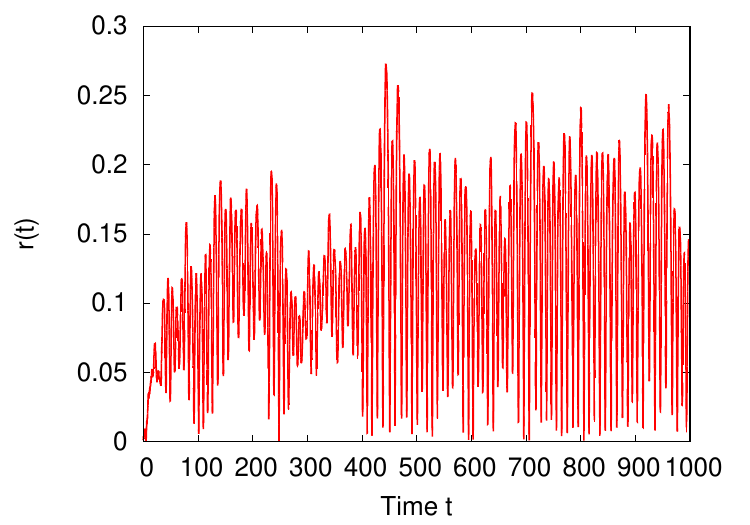}
\caption{For the model~\eqref{eq:eom}, the figure shows for the case studied in Section~\ref{sec:choice2} of the particular dependence of the diffusion
coefficient $D$ on $\omega$ the order parameter $r$ versus time and for $\gamma=1.5,K=2.83$, such that $K_i < K < K_c$. Upper panel: initial state has all
the $\theta$'s having the same value. Lower panel: initial state has the $\theta$'s uniformly and independently distributed
in $[-\pi,\pi]$. The frequency distribution is a Gaussian with zero mean and unit variance. The data are obtained from numerical integration of the equations
of motion~\eqref{eq:eom}, with the number of oscillators $N=10^5$. The figure suggests bistability, i.e., coexistence of stable periodic and synchronized states.}
\label{fig:simulation-CaseB}
\end{figure}

The situation for $\gamma = 1.5$ is again different and probably more unusual. We see from from panel (b) of
Fig.~\ref{fig:CaseB-r-vs-K} that now the instability threshold $K_i$ of the incoherent state does not coincide with the bifurcation threshold
$K_c$ ($K_i \approx 2.784$, $K_c \approx 3.182$); however, contrary to what we saw for larger values of $\gamma$,
we have $K_i$ larger than the instability threshold $K_i^{(\mathrm{sync})} \approx 2.625$ of the synchronized state. At
first, one may guess that, similar
to what happens for $\gamma = 1.0$, there is a hysteresis behavior in the range  $K_i^{(\mathrm{sync})} < K < K_i$. However,
 numerical integration results show that by adiabatically increasing $K$ beyond $K_i$, the system does not jump
to the synchronized state, but it settles into a nonstationary periodic state. We have not determined the
value of $K$ between $K_i$ and $K_c$ at which this nonstationary asymptotic state ceases to exist; however, in our
opinion, the interesting thing is that there is a range of $K$ for which we have another kind of bistability, namely, that
between a nonstationary periodic state and a synchronized state. To show this fact, in Fig.~\ref{fig:simulation-CaseB},
we plot $r(t)$ obtained from two simulations at $K=2.83$, a value between $K_i$ and $K_c$. In one simulation, the initial
configuration is given by all phases equal to one another, and the system goes to the synchronized stationary state; in the other
simulation, the initial configuration has the phases uniformly and independently distributed in $[-\pi,\pi]$, and we see that the system relaxes to a nonstationary state.
Thus, also for this case with $\gamma=1.5$ the value $K_i$
marks a Hopf bifurcation. However, contrary to the cases with higher values of $\gamma$, the periodic state for $K>K_i$
coexists, for a certain $K$ range, with a stationary synchronized state.

\section{Discussion and conclusions}
\label{eq:conclusions}

In this work, we have studied the synchronization properties of a system of Kuramoto oscillators subjected to a noise strength that is not the same for all the
oscillators. This is at variance with the usual statistical mechanical treatment of
many-body systems subjected to thermal noise. In this latter case, all the components of the system being subjected to the same
temperature, the strength of the noise, in an analysis through Langevin equations or a Fokker-Planck equation, is
uniform. We emphasize that in our studied system, the noise represents not only the thermal fluctuations induced by the interaction
with a heat bath (i.e., the external environment), but also the variability of the intrinsic frequencies of the
individual components.

Our study has revealed an interesting feature about the role of noise in a system of interacting oscillators. In fact, it has
generally been found in systems with homogeneous/uniform noise that noise tends to simplify the overall behavior. This is not so
much evident when the frequency distribution $g(\omega)$ is symmetric and unimodal (in this case, the noise simply causes
from a practical point of view a shift of the curve $r(K)$ versus $K$ along the $K$-direction), but which becomes clearer with more
structured forms of $g(\omega)$. For example, the rich phase diagram found with a symmetric and bimodal $g(\omega)$ without
noise~\cite{Martens2009}, while being qualitatively of the same structure at small noise, is progressively simplified with 
increasing noise levels~\cite{Campa2020}. This can be understood by noting that increasing noise tends to smoothen the
structure of $g(\omega)$ (in particular, it tends to merge the two peaks of bimodal $g(\omega)$), making the effects of this distribution similar
to those of a unimodal one. Instead, we have seen in the current work that the introduction of a noise with inhomogeneous strength/intensity, in particular,
a noise which is more intense for small frequencies and less so for large frequencies, can transform the simple behavior
occurring in a system of oscillators with symmetric and unimodal $g(\omega)$ into a richer structure, including
(nonequilibrium) first-order phase transitions, hysteresis behavior, nonstationary asymptotic states. 

Trying to rationalize
qualitatively our observed behavior while going beyond the results of our analytic computation and numerical
simulations, one can argue as follows. As a first thing, one may note that in the homogeneous-noise case, the stability of
the incoherent state is increased by increasing noise, while the existence and stability of the synchronized state
is a much more difficult affair. In fact, noise facilitates the spreading of oscillator phases over the range $[-\pi,\pi]$. This is why by adding noise to a
system with symmetric and unimodal $g(\omega)$, the curve of $r(K)$ versus $K$ is progressively shifted to larger $K$ while maintaining the same simple
form. To give a more quantitative information on this, we can refer to Fig.~\ref{fig:r-vs-K-bare-model} that shows
the curve $r(K)$ for the homogeneous-noise strength equal to $D=1.0$: in this case, the bifurcation threshold is
$K_c =K_i \approx 3.050$; if we study the same system with $D=0$, the curve $r(K)$ versus $K$ will be similar, but with the
bifurcation threshold at $K_c =K_i \approx 1.596$. Let us now compare the values of $K_i$ and $K_i^{(\mathrm{sync})}$ for the
following four cases: $D=0$, case (i) of Section~\ref{sec:choice1}, case $\gamma = 4.0$ of Section~\ref{sec:choice2}, and the homogeneous-noise case with
$D=1.0$; we stress that in the last three cases, the noise strength at small frequencies is always
$D\to 1$. The values of $K_i$ are: $K_i=1.596$ for $D=0$, $K_i=1.697$ for case (i) of Section~\ref{sec:choice1},
$K_i=1.843$ for the case $\gamma=4.0$ of Section~\ref{sec:choice2}, and $K_i=3.050$ for the homogeneous-noise case with $D=1.0$. We note that the second
and the third value are much closer to the first one than to the last one. One infers from this that the stability of
the incoherent state is not increased much when the noise is mainly concentrated on small frequencies. On the other hand,
let us consider the values of $K_i^{(\mathrm{sync})}$, which for the first and the last of the mentioned four cases coincide with $K_i$:
$K_i^{(\mathrm{sync})}=1.596$ for $D=0$, $K_i^{(\mathrm{sync})}=1.995$ for case (i) of Section~\ref{sec:choice1},
$K_i^{(\mathrm{sync})}=2.186$ for case $\gamma=4.0$ of Section~\ref{sec:choice2}, and $K_i^{(\mathrm{sync})}=3.050$ for the homogeneous-noise case with
$D=1.0$. The values of the second and the third case suggest that the noise, although not able to stabilize much the incoherent
state, is nevertheless sufficient to prevent the existence, for a certain $K$ range beyond $K_i$, of the synchronized state.
This can be explained by the fact that the existence of the synchronized state requires that oscillators of all frequencies, in
particular, the small ones, are not too much affected by the noise. This is only a heuristic argument, which we believe
grasps the real physical reason of the phenomenon that we have found.

In this study, we have considered the noise strength $D$ to be a function of frequency $\omega$, so that oscillators with the same $\omega$ share
the same noise strength. Mathematically, this can be represented by a joint quenched frequency-noise distribution
of the form $F(\omega,D)=g(\omega)\delta(D-D(\omega))$. A more general situation would be the one with a general form
for $F(\omega,D)$, including the case of independent frequencies and noise, i.e., $F(\omega,D)=g(\omega)P(D)$. We
have not studied this more general case, which would be more elaborate from the analytical point of view, but which
above all would require simulations with much larger system sizes to have statistically-meaningful results. On the other
hand, if we consider our starting motivation for introducing inhomogeneous noise, i.e., different intrinsic
frequency variability among the oscillators, it is sensible to assume that oscillators with the same intrinsic frequency
are subjected to the same noise. It is also sensible to assume, as for the cases where we have found the interesting
and complex behavior of our system, that oscillators with small frequencies are more prone to be affected by the
environment~\cite{note2}, and so in the model, one has to take for them a larger noise strength.

Even staying within our choice of $F(\omega,D)=g(\omega)\delta(D-D(\omega))$, it is clear
that each choice of the form of $D(\omega)$ would produce quantitatively a different structure for the possible
asymptotic states (which, as previously noted, can be represented for a given $g(\omega)$ in a phase diagram with
coordinates given by $K$ and the parameters of $D(\omega)$). It is then more useful to try to determine the qualitative
features of these different phase diagrams that would be preserved going from one form of $D(\omega)$ to another one, at least
within a class of functions with a given property, i.e., $D(\omega)$ being a decreasing function of $|\omega|$ as we have mostly
considered in this work. In this first work, we have studied only a couple of different forms of the function $D(\omega)$, the step function
in Section~\ref{sec:choice1} (case (i)) and the smooth function of the form analyzed in Section~\ref{sec:choice2}. We feel that
the qualitative features that we found for these choices of $D(\omega)$ would be present in most cases in which the
function $D(\omega)$ has the mentioned property, but clearly this assumption should be supported by an extensive
investigation that is left to future studies. 

\section*{Acknowledgements}

SG acknowledges support from the Science and Engineering Research Board (SERB), India under SERB-CRG Grant CRG/2020/000596. SG also thanks ICTP, Trieste, Italy, for
support under its Regular Associateship scheme. AC and SG thank the hospitality of SISSA and ICTP, Trieste, Italy for support and warm hospitality during July 2023
when this work was being written up. 

\appendix

\section{The bifurcation threshold}
\label{sec:app1}

We begin by writing explicitly Eq.~\eqref{kthreshold_step} determining the bifurcation threshold $K_c$:
\begin{eqnarray}
\label{kbif1}
K_c &=& \sqrt{8\pi}\left[ \int_{|\omega|\le \omega_0} \mathrm{d}\omega~ e^{-\frac{\omega^2}{2}} \frac{D_1}{\omega^2 + D_1^2}
\right. \nonumber \\
&+& \left. \int_{|\omega|\ge \omega_0}  \mathrm{d}\omega~ e^{-\frac{\omega^2}{2}} \frac{D_2}{\omega^2 + D_2^2} \right]^{-1}.
\end{eqnarray}
We see that to compute this threshold, it is necessary to evaluate an integral of the type:
\begin{equation}
\label{iabalph}
I(a,b,\alpha) \equiv \int_{-\alpha}^{+\alpha} \mathrm{d} x \, \frac{e^{-ax^2}}{x^2+b^2}.
\end{equation}
The usual case in which $D$ is a constant for all $\omega$'s requires the value of the above integral for $\alpha=\infty$, and we know that
\begin{equation}
\label{iabinf}
I(a,b,\infty) = \int_{-\infty}^{+\infty} \mathrm{d} x \, \frac{e^{-ax^2}}{x^2+b^2} = \frac{\pi}{b}\mathrm{erfc}(\sqrt{ab^2})e^{ab^2},
\end{equation}
where $\mathrm{erfc}(x)$ is the complementary error function, defined in terms of the error function $\mathrm{erf}(x)$ as
\begin{equation}
\label{erferfc}
\mathrm{erf}(x) \equiv \frac{2}{\sqrt{\pi}}\int_0^x \mathrm{d} t \, e^{-t^2};~~~~
\mathrm{erfc}(x) \equiv 1 -\mathrm{erf}(x).
\end{equation}

The evaluation of the integral \eqref{iabalph} can be performed with a procedure similar to that for the computation of \eqref{iabinf}.
Computing the partial derivative of $I(a,b,\alpha)$ with respect to $a$, we easily obtain
\begin{eqnarray}
\label{parder}
\frac{\partial}{\partial a}I(a,b,\alpha) &=& b^2 I(a,b,\alpha) - \int_{-\alpha}^{+\alpha} \mathrm{d} x \, e^{-ax^2}\nonumber \\
&=& b^2 I(a,b,\alpha) -\sqrt{\frac{\pi}{a}} \mathrm{erf}(\alpha \sqrt{a}).
\end{eqnarray}
The above is a differential equation for $I(a,b,\alpha)$, whose solution, knowing the value for $a=0$ given by
\begin{equation}
\label{initval}
I(0,b,\alpha) = \int_{-\alpha}^{+\alpha} \mathrm{d} x \, \frac{1}{x^2+b^2} = \frac{2}{b} \arctan \left(\frac{\alpha}{b}\right),
\end{equation}
is
\begin{eqnarray}
\label{soliab}
&&I(a,b,\alpha)\nonumber \\
&&=\left\{ \frac{2}{b} \arctan \left(\frac{\alpha}{b}\right) -2\sqrt{\pi} \int_0^{\sqrt{a}} \mathrm{d} t \, e^{-b^2t^2}
\mathrm{erf}\left(\alpha t\right)\right\}e^{ab^2}.\nonumber \\
\end{eqnarray}

To proceed, let us define
\begin{equation}
\label{interf}
J(\alpha,b,a) \equiv \int_0^{\sqrt{a}} \mathrm{d} t \, e^{-b^2t^2}
\mathrm{erf}\left(\alpha t\right).
\end{equation}
We note that this integral can be expressed in closed form when $\alpha=b$. In fact, the definition of the
error function implies that $(\mathrm{d}/\mathrm{d} t)\mathrm{erf}(bt) = (2b)/\sqrt{\pi}~e^{-b^2t^2}$, which in turn
implies that the integrand in Eq.~\eqref{interf} is $\frac{\sqrt{\pi}}{4b} \frac{\mathrm{d}}{\mathrm{d} t} \left[\mathrm{erf}^2(bt)\right]$, so
that $J(b,b,a) = \frac{\sqrt{\pi}}{4b} \mathrm{erf}^2(b\sqrt{a})$. For $\alpha \ne b$, we can write, using 
$(\mathrm{d}/\mathrm{d} t)\mathrm{erf}(t) = 2/\sqrt{\pi}~e^{-t^2}$, that
\begin{eqnarray}
\label{interfg}
J(\alpha,b,a)  &=& \frac{\sqrt{\pi}}{4b} \mathrm{erf}^2(b\sqrt{a})\nonumber \\
&&+\frac{2}{b\sqrt{\pi}} \int _0^{b\sqrt{a}} \mathrm{d} v \, \int_v^{\frac{\alpha}{b}v} \mathrm{d} u \, e^{-(v^2+u^2)}.
\end{eqnarray}
This is valid for both $\alpha > b$ and $\alpha < b$. In the double integral on the right hand side, we can go to polar
coordinates, defined by $v \equiv w\cos \phi$, $u \equiv w\sin \phi$, to obtain
\begin{eqnarray}
\label{double}
&&\int _0^{b\sqrt{a}} \mathrm{d} v \, \int_v^{\frac{\alpha}{b}v} \mathrm{d} u \, e^{-(v^2+u^2)}\nonumber \\
&&= \frac{1}{2}\int_{\frac{\pi}{4}}^{\arctan \left(\frac{\alpha}{b}\right)} \mathrm{d} \phi \,
\left( 1 - e^{-r^2(\phi)}\right).
\end{eqnarray}
From trigonometric considerations, we have $r^2(\phi)= ab^2\left(1+\tan^2\phi\right) = \frac{ab^2}{\cos^2\phi}$. We therefore
have:
\begin{eqnarray}
\label{intj}
&&J(\alpha,b,a) = \frac{\sqrt{\pi}}{4b} \mathrm{erf}^2(b\sqrt{a}) \nonumber \\
&&+ \frac{1}{b\sqrt{\pi}}
\int_{\frac{\pi}{4}}^{\arctan \left(\frac{\alpha}{b}\right)} \mathrm{d} \phi \, \left( 1 - e^{-\frac{ab^2}{\cos^2\phi}}\right).
\end{eqnarray}
We can further transform this expression to get rid of the square of the error function. For this purpose, we note that
\begin{equation}
\label{jinf}
J(\infty,b,a) = \int_0^{\sqrt{a}} \mathrm{d} t \, e^{-b^2t^2} = \frac{\sqrt{\pi}}{2b}\mathrm{erf}(b\sqrt{a}).
\end{equation}
This implies that
\begin{equation}
\label{jsimp}
\frac{1}{b\sqrt{\pi}} \int_{\frac{\pi}{4}}^{\frac{\pi}{2}} \mathrm{d} \phi \, \left( 1 - e^{-\frac{ab^2}{\cos^2\phi}}\right)
= \frac{\sqrt{\pi}}{2b}\mathrm{erf}(b\sqrt{a}) - \frac{\sqrt{\pi}}{4b} \mathrm{erf}^2(b\sqrt{a}).
\end{equation}
Then, Eq.~\eqref{intj} can be transformed to
\begin{eqnarray}
\label{intj1}
&&J(\alpha,b,a)\nonumber \\
 &&= \frac{\sqrt{\pi}}{4b} \mathrm{erf}^2(b\sqrt{a}) + \frac{1}{b\sqrt{\pi}}
\int_{\frac{\pi}{4}}^{\frac{\pi}{2}} \mathrm{d} \phi \, \left( 1 - e^{-\frac{ab^2}{\cos^2\phi}}\right)\nonumber \\
&&- \frac{1}{b\sqrt{\pi}} \int_{\arctan \left(\frac{\alpha}{b}\right)}^{\frac{\pi}{2}} \mathrm{d} \phi \,
\left( 1 - e^{-\frac{ab^2}{\cos^2\phi}}\right) \nonumber \\
&&= - \frac{\sqrt{\pi}}{2b}\mathrm{erfc}(b\sqrt{a}) +
\frac{1}{b\sqrt{\pi}} \arctan \left(\frac{\alpha}{b}\right) \nonumber \\
&&+ \frac{1}{b\sqrt{\pi}}
\int_{\arctan \left(\frac{\alpha}{b}\right)}^{\frac{\pi}{2}} \mathrm{d} \phi \,
e^{-\frac{ab^2}{\cos^2\phi}}.
\end{eqnarray}

Now, inserting the above result on the right hand side of Eq.~\eqref{soliab}, we obtain for $I(a,b,\alpha)$ the expression:
\begin{equation}
\label{soliab2}
I(a,b,\alpha) = \frac{\pi}{b}\mathrm{erfc}(b\sqrt{a})e^{ab^2} -\frac{2}{b}
\int_{\arctan \left(\frac{\alpha}{b}\right)}^{\frac{\pi}{2}} \mathrm{d} \phi \, e^{-ab^2\tan^2\phi}.
\end{equation}
We note that for $\alpha = \infty$, we obtain the correct expression \eqref{iabinf}. Using Eq.~\eqref{soliab2} in
Eq.~\eqref{kbif1}, we finally have
\begin{eqnarray}
\label{kbiffin}
K_c &=& \sqrt{\frac{8}{\pi}} \Big[ D_1 I\left(\frac{1}{2},D_1,\alpha\right) +D_2 I\left(\frac{1}{2},D_2,\infty\right) \nonumber \\
&&- D_2 I\left(\frac{1}{2},D_2,\alpha\right) \Big]^{-1} \nonumber \\
&=& \sqrt{\frac{8}{\pi}} \Big[ \mathrm{erfc}\left(\frac{D_1}{\sqrt{2}}\right)e^{\frac{D_1^2}{2}}
-\frac{2}{\pi} \int_{\arctan \left(\frac{\alpha}{D_1}\right)}^{\frac{\pi}{2}} \mathrm{d} \phi \, e^{-\frac{D_1^2}{2}\tan^2\phi}\nonumber \\
&&+\frac{2}{\pi} \int_{\arctan \left(\frac{\alpha}{D_2}\right)}^{\frac{\pi}{2}} \mathrm{d} \phi \, e^{-\frac{D_2^2}{2}\tan^2\phi}
\Big]^{-1},
\end{eqnarray}
which is Eq.~\eqref{kbiffin-main} of the main text with the substitution $\alpha=\omega_0$. 
Since from the definition~\eqref{iabalph} we have that $I(a,b,0)=0$, from Eq.~\eqref{soliab2}, we obtain
\begin{equation}
\label{soliab0}
\mathrm{erfc}(b\sqrt{a})e^{ab^2} = \frac{2}{\pi}
\int_{0}^{\frac{\pi}{2}} \mathrm{d} \phi \, e^{-ab^2\tan^2\phi}.
\end{equation}
Then, we see from Eq.~\eqref{kbiffin} that, as desired, the limit $\alpha \to \infty$ yields the threshold for the case of constant
$D$ equal to $D_1$, while the limit $\alpha \to 0$ yields the threshold for the case of constant $D$ equal to $D_2$. We also
obtain the threshold for constant $D$ when $D_1=D_2=D$.

\section{The numerical solution of Eq.~\eqref{eq:steady2-1} with $f_\mathrm{st}$ given by Eq.~\eqref{eq:solution-D}}
\label{sec:app2}

We discuss here the numerical solution of Eq.~\eqref{eq:steady2-1} with $f_\mathrm{st}$ given by Eq.~\eqref{eq:solution-D}. Let us then start
from the stationary Fokker-Planck equation~\eqref{eq:steady1-1} with $D=D(\omega)$, given in the equivalent form by
\begin{equation}
\label{FPcurrent_sol_b}
\left( \omega -Kr\sin\theta\right) f_\mathrm{st}(\theta,\omega)
- D(\omega) \frac{\partial}{\partial \theta}f_\mathrm{st}(\theta,\omega) = S(\omega),
\end{equation}
where $S(\omega)$ is the constant and uniform probability current in the stationary state, and $f_\mathrm{st}(\theta,\omega)$ is the normalized stationary
distribution function given by Eq.~\eqref{eq:solution-D}. We need to use the expression
\begin{equation}
\label{expectipt}
\int_0^{2\pi} \mathrm{d} \theta \, e^{\mathrm{i} p\theta} f_\mathrm{st}(\theta,\omega) = \langle e^{\mathrm{i} p\theta} \rangle (\omega;r)
\end{equation}
of the expectation value (which depends on $\omega$ and $r$) of the function $e^{\mathrm{i} p\theta}$ with respect to the distribution
$f_\mathrm{st}(\theta,\omega)$, where $p$ is any positive integer. Then, multiplying Eq.~\eqref{FPcurrent_sol_b} by $e^{\mathrm{i} p\theta}$ and
integrating over $\theta$, we obtain for any given $\omega$ the following
system of equations:
\begin{eqnarray}
\label{FPcurrent_p1}
&&\left( \omega +\mathrm{i} D(\omega)\right)\langle e^{\mathrm{i} \theta}\rangle (\omega;r) +\mathrm{i}
\frac{Kr}{2}\langle e^{\mathrm{i} 2\theta}\rangle (\omega;r)= \mathrm{i} \frac{Kr}{2}, \nonumber \\
\label{FPcurrent_pgen}
&&\left( \omega +\mathrm{i} pD(\omega)\right)\langle e^{\mathrm{i} p\theta}\rangle (\omega;r) +\mathrm{i}
\frac{Kr}{2}\langle e^{\mathrm{i} (p+1)\theta}\rangle (\omega;r) \nonumber \\
&&-\mathrm{i} \frac{Kr}{2}\langle e^{\mathrm{i} (p-1)\theta}\rangle (\omega;r)= 0;\,\,\,\,\,\,\,\,\,\,\,\,\,\,\,\,\, p=2,3,\dots
\end{eqnarray}
Solving the above system of equations by truncating them at a large value of $p$, one obtains $\langle e^{\mathrm{i} p\theta}\rangle(\omega;r)$
for each $p$. Using the result for $p=1$ in the self-consistent equation~\eqref{eq:steady2-1}, which   may be written as
\begin{equation}
\label{selffourier}
r = \int \mathrm{d} \omega~g(\omega) {\rm Re}\left[\langle
e^{\mathrm{i} \theta}\rangle (\omega;r)\right],
\end{equation}
one has to next solve numerically the above self-consistent equation for $r$, obtaining finally the desired dependence of $r$ on $K$.



\begin{thebibliography}{99}
\bibitem{note2}
Oscillators with a small frequency could be more affected, from the environment and from the internal processes, not only in the dynamics of their phase, but also
in their amplitude. Since the Kuramoto model uses a phase-only approximation, this larger sensitivity could be modeled by a larger noise strength acting on the phase.
\end{thebibliography}
\end{document}